# A Comprehensive Approach to Diagnosing Temporomandibular Joint Diseases: AI-driven TMD Diagnostic System

Y. Gu[a+], C.T. Kong[b+], D.D Zhang[c], Y.J Bai[d], J.K.H. Tsoi[a], Hua Huang[c], Y.Q. Deng[c*] , Y.M Zhu[e]

a. Faculty of Dentistry, The University of Hong Kong, Prince Philip Dental Hospital, 34 Hospital Road, Sai Ying Pun, Hong Kong

b. Faculty of Engineering, The University of Hong Kong, Haking Wong Building, Pokfulam Road, Hong Kong

c. Department of Stomatology, Shenzhen University General Hospital, 1098 Xueyuan Road, Shenzhen City, Guangdong Province, the People's Republic of China

d. The First Clinical Medical College of Zhejiang Chinese Medical University, Hangzhou, the People's Republic of China

e. Department of Oral and Maxillofacial Surgery, Shenzhen Dental Hospital,70# Gui Yuan Bei Road, Luo Hu District, Shenzhen City, Guang Dong Province, 518001, the People's Republic of China

[+] both authors contributed equally to the manuscript.

*Abstract*— **AI-driven TMD diagnostic system uses AI segmentation method to diagnose Temporomandibular Joint Disorders (TMD). By using segmentation, three important parts: temporal bone, temporomandibular joint (TMJ) disc and the condyle can be identified. The location and the size of each segment are used as the basic information to determine if the patient has a high chance of having Temporomandibular Joint Disorders (TMD).**

## I. Introduction

Magentic Resonance Imaging (MRI) is one of the important methods for diagnosing Temporomandibular Joint Disorders (TMD). However, the usefulness of MRI remains challenged in the accurate and efficient diagnosis of diseases from the result images. Artificial Intelligence (AI) methods have shown great promise in analyzing medical images and improving diagnostic outcomes.

Meanwhile, interpreting MRI results for TMD diagnoses remains challenging due to the complex anatomical structures, variations, and potential overlap with other conditions. The recent development of Artificial Intelligence (AI) methods, particularly segmentation [1], has demonstrated significant potential in the analysis of medical images, leading to improved diagnostic outcomes in various medical fields.

## II. Literature Review

In the context of TMD diagnosis, the application of AI has shown promising results in the literature. The study by Bai G et al.[2] presented an innovative attention-based multimodal deep learning AI system, called TMJ MRI-Net system, which employed attention-based multimodal deep learning technology to overcome the challenges associated with the unidirectional spatial sparse MRI data for diagnosing temporomandibular joint (TMJ) disc displacement field. This novel system effectively bypasses the limitations of conventional 3D deep learning models, such as 3D convolution or reconstructed mesh, that were ill-suited for the unique spatial sparsity of TMJ MRI data. By using 8 layers of MRI images, the tip of condyle was identified. The images were cropped with the tip of condyle as the center. Then, the feature network was used to extract the feature and output the result. However, this study was not without limitations. The current TMJ MRI-Net system exhibited difficulty in predicting the internal and external displacement of the articular disc based on 8 TMJ MRI images, with a diagnostic sensitivity of only 78% in the machine learning testing set [2]. This result suggests that the AI framework may face challenges in forming its three-dimensional space structure prediction, implying the need for the development of a new three-dimensional perspective learning system.

By employing helpful techniques, radiologists will be able to improve the speed and accuracy of TMD diagnoses, ultimately leading to better patient outcomes. However, there are still existing technical barriers to be overcome, such as implementation, ethical considerations, and the need for extensive training data.

For our approach, as the condyle (shown in figure 1) was not difficult to find in the MRI image, cropping image was done manually. Afterward, by using segmentation to identify the temporal bone, temporomandibular joint (TMJ) disc and condyle, the position and size of each segment can be used for diagnosis.

### III. BACKGROUND

Temporomandibular Joint Disorder (TMD) is a medical term denoting dysfunction of the temporomandibular joint and its associated structures, characterized by clicking, pain, and functional limitations within the maxillofacial region.[3] Patients with untreated TMD may endure chronic headaches, earaches, and eating difficulties over the long term, thereby diminishing their quality of life. Early diagnosis of TMD is crucial as it enables intervention at early stage, potentially preventing the disorder from advancing to bony resorption and minimizing the risk of developing chronic symptoms.[4] [5] Following early intervention, treatment can be administered using less invasive methods, potentially circumventing the need for surgery or long-term pharmacological dependence.[6]

The objective of this study is to develop an AI-based method for analyzing TMD MRI images. In this paper, we will discuss our research methodology, including data acquisition, preprocessing, model training, and the ultimate AI-driven TMD diagnostic system, as well as the potential implications of our findings for the clinical treatment of TMD. Also, we will discuss our research methods including custom-python coding, data training and outcome interpretation by employing segmentation techniques to improve the speed of disease diagnosis, which would be both convenient for clinicians and patients. The MRI scan below shows the side view of a skull. The important part of the MRI scan is at the tip of condyle.

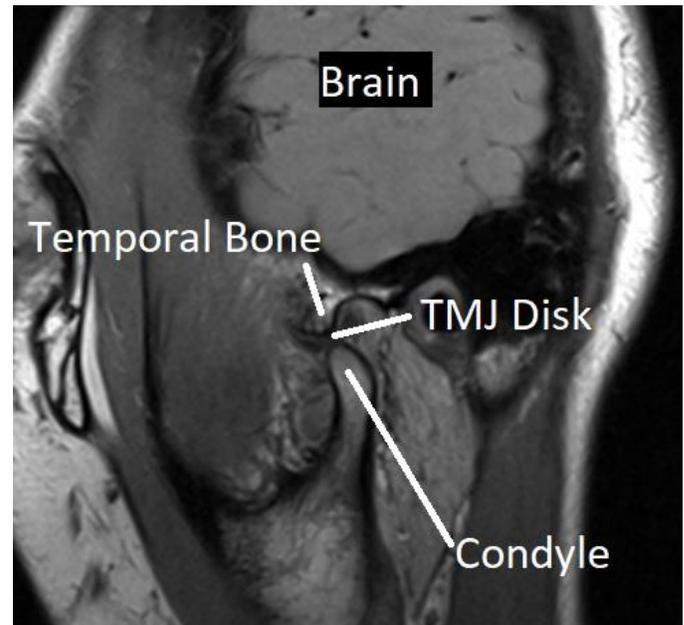

*Fig 1. MRI scan of a skull (side view).*

Several different methods were tried before our AI-driven TMD diagnostic system was finalized. Before machine learning, Computer vision was used to extract the contour of the condyle. In the picture below, the tip and the two sides of the condyle have been analyzed to determine whether the condyle shape was normal. However, this method could not diagnose accurately, as different MRI slices of the same patient may yield different results.

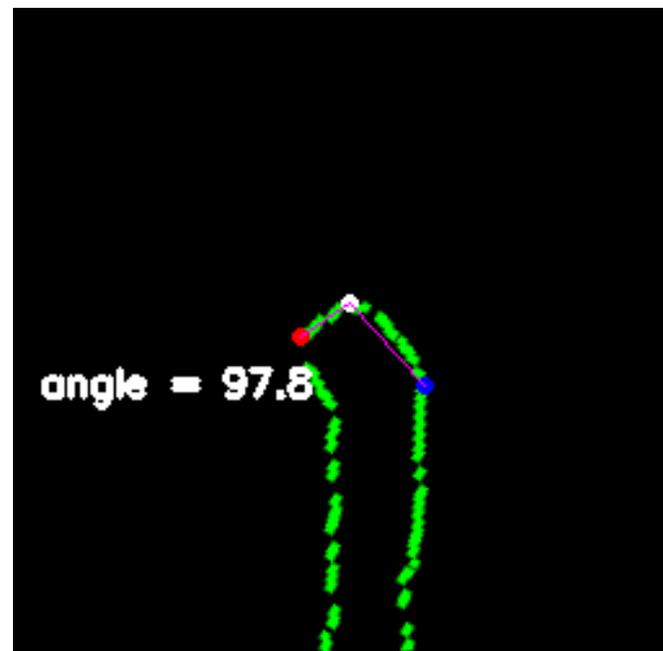

*Fig 2. Using OpenCV to extract the contour of the condyle. The red dot is the far-left point, the white dot is the top point, and the blue point is the far-right point. The angle value is the acute angle between two lines.*

Before using segmentation, a simpler method called object detection was employed [7]. This method required less labelling time by drawing boxes only. However, the shape of the condyle and disc were irregular, and the box area was too big and covered many unused pixels area. And we tried to have two labels of "TMJ_disc" and "TMJ_disc_abnormal" label which produced low accuracy and inconclusive results.

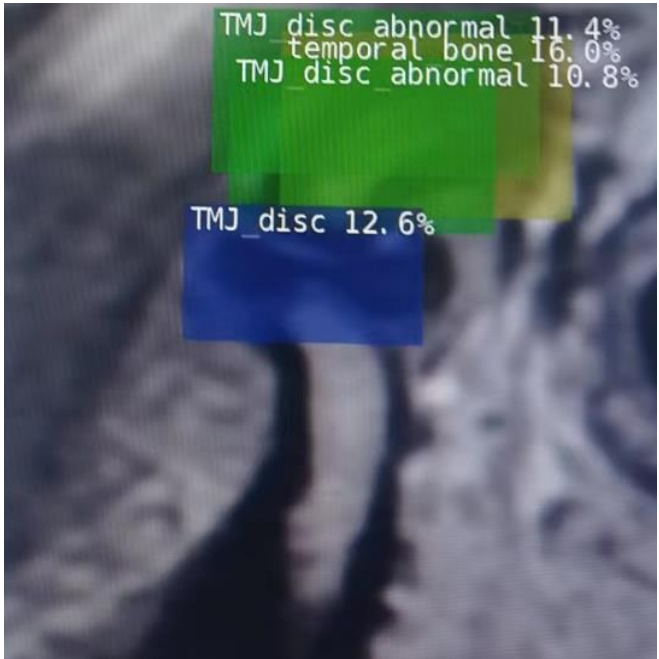

*Fig3. Object detection which produced a low accuracy result*

As the shape was irregular, segmentation was used [1]. For the segmentation, we initially struggled with choosing the proper number of labels, and determining which parts in the MRI were important to be analyzed. We finalized with 3 labels: temporal bone, TMJ disc, and condyle.

For the pretrained model, the fully connected network [8], fcn_resnet18 [1] was used for segmentation. However, the model did not give sufficient edge details. Therefore, we used fcn_resnet101 instead.

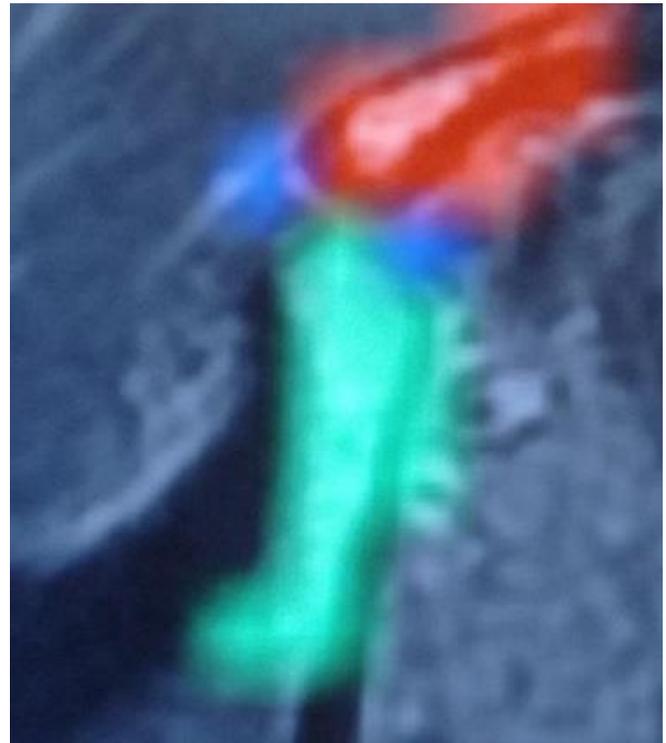

*Fig4. Segmentation based on fcn_resnet18.*

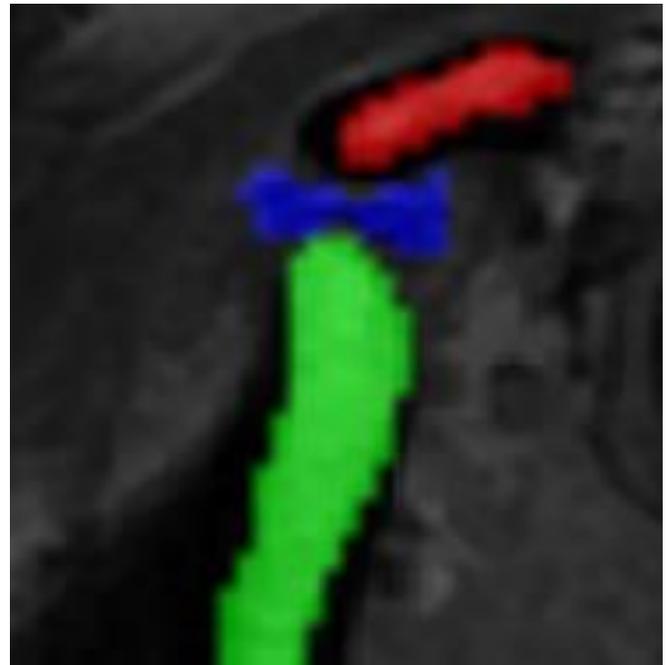

*Fig5 Segmentation based on fcn_resnet101.*

In the initial testing, the entire MRI image was analyzed. However, we found out the image size was too big. Many useless structures were included that affected the segmentation training. Cropping of the MRI image was necessary.

For the inference, we tried to perform classification based on the RGB color area output from the segmentation. However, the result was not acceptable. The low accuracy may have been due to the small dataset size. Finally, a simpler decision tree method based on the geometrical relationship of each segment was used to determine whether the patient was normal.

## IV. SCOPE OF WORK

**Data Source**

This study was approved by the Research Ethics Committee of the Shenzhen General Hospital (Protocol Number KYLL-20221217A). The data training and outcome inference procedures were conducted at the University of Hong Kong.

MRI examinations were performed on a 3T MRI scanner (Magnetom Skyra, Siemens Healthineers). MRI sequences included T1-weighted, T2-weighted, and proton density-weighted images in the sagittal and coronal planes, while MRI scans were performed with 2 mm slice thickness. MRI images were anonymized and saved in png format for further analysis.

A retrospective database comprising the image records of 140 clinical cases was created for AI segmentation training. The dataset consisted of 62 abnormal and 78 normal images, which had been pre-validated by expert radiologists. For our training, 80% of dataset was the training set, 10% was the validation set and 10% was the testing set.

**AI-driven TMD Diagnostic System**

The development of the AI-driven TMD diagnostic system consisted of two phases: Training and Inferencing. Training was used to train the segmentation model. Inferencing was used to diagnose if the patient has TMD.

## V. TRAINING

Below is the Training block diagram.

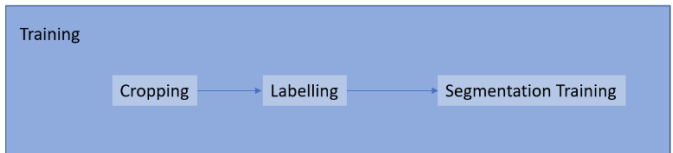

*Fig6. The Training block diagram*

**Cropping** - crop out the region of interest.
MRI image below shows the cross section of the skull, and the region of interest is around the condyle. A section of 200x200 pixels with the center at the tip of the condyle was cropped for further analysis.

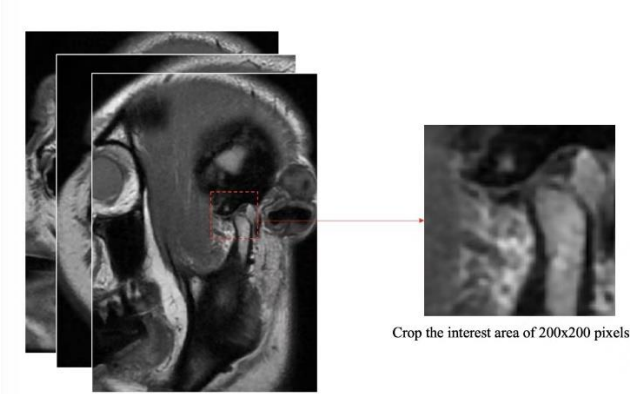

*Fig7. Cropping 200x200 pixels around condyle*

**Labelling** – draw polygons for different parts.
Three labels were used for segmentation labelling: temporal bone, TMJ discs and condyle. An open-source software called Labelme was used for drawing the polygon of each segment on each MRI image [9].

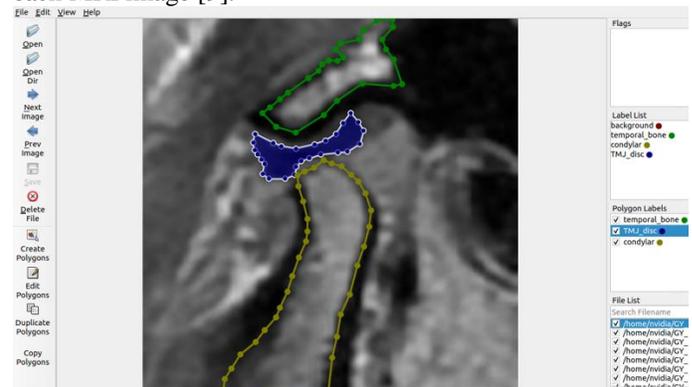

*Fig8. LabelMe User Interface for Labelling*

**Segmentation Training** – train the segmentation model.
The output file format was Pascal VOC2012 [10]. After the dataset set was collected, the transfer learning based on fcn resnet101 [11] was used for training. After the training, the output file "model_best.pth" was converted to the onnx model. The training section was completed. The onnx model [12] was the machine learning model for inference in the section below. [13]

## VI. INFERENCE

Below is the Inference block diagram.

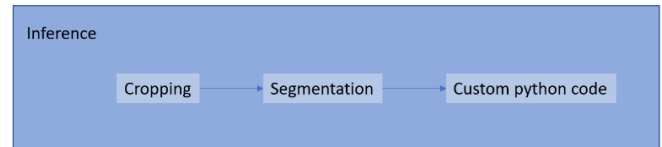

*Fig9. The Inference block diagram*

**Cropping** - crop out the region of interest.
Same as the training, the MRI scan was cropped with the tip of condyle in the center of the picture. The size of the cropped picture was 200 x 200 pixels.

**Segmentation** – identify the area of each label.
The onnx model from the segmentation training was used to identify the segment of each label. Red color indicated the temporal bone, blue indicated the TMJ disc and green indicated the condyle.

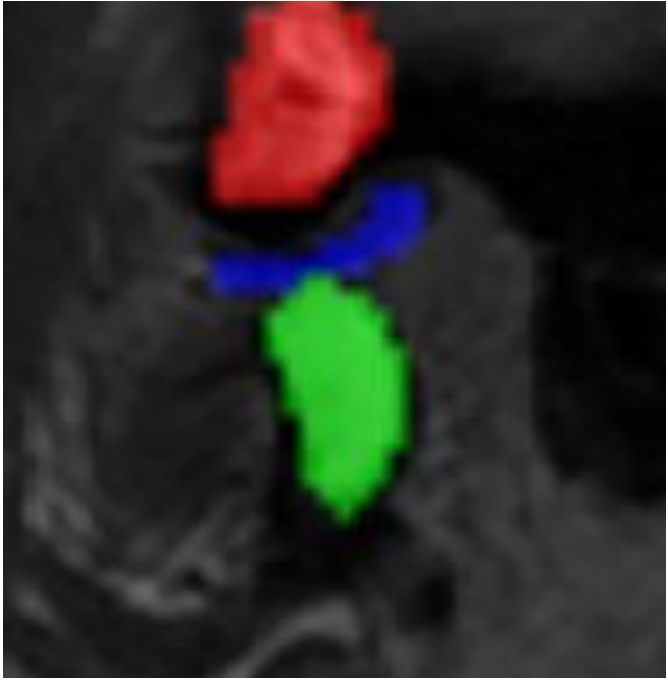

*Fig10. The overlay of segmentation on MRI scan. Red is the temporal bone, blue is the TMJ disc and green is the condyle.*

**Custom python code** – determine if the patient was normal or abnormal.

The area and the center point of each segment were calculated. By using the geometrical relationship of this information, the custom-made Python program used a decision tree to determine if the patient was normal or abnormal.

*There are five cases to determine if it is abnormal.*
*R: Temporal Bone*
*B: TMJ Disk*
*G: Condyle*
*BRtolerance, BGtolerance, Bmintolerance, Bmaxtolerance, BminHeight, Gmintolerance: preset value*
*Case1: If neither TMJ Disk nor Condyle can be seen.*
*Case 2: For x axis, (R(center)-B(center)) + BRtolerance < 0*
*Case 3: For x axis, (B(center) -G(tipX)) + BGtolerance < 0*
*Case 4: For x axis, B(width) < Bmintolerance or (B(width) > Bmaxtolerance and B(height) < BminHeight)*
*Case 5: For y-axis, B(center) > (GminY+Gmintolerance)*

*Case 1*
*If the TMJ disc or Condyle cannot been identified, the image is considered to be abnormal.*
*Case 2*
*If the TMJ disc center point is on the left side of the temporal bone center point beyond a preset parameter (BRtolerance), the image is considered to be abnormal.*
*Case 3*
*If the TMJ disc center point is on the left side of tip of Condyle beyond a preset parameter (BGtolerance), the image is considered to be abnormal.*
*Case 4*
*If the TMJ disc width is smaller than the preset parameter (Bmintolerance), or (the TMJ disc width is bigger than the preset parameter (Bmaxtolerance) and the TMJ height is smaller than the preset parameter (BminHeight)), image is considered to be abnormal.*

*Case 5*
*If the TMJ disc center is lower than the tip of Condyle beyond a preset parameter (Gmintolerance), the image is considered to be abnormal.*

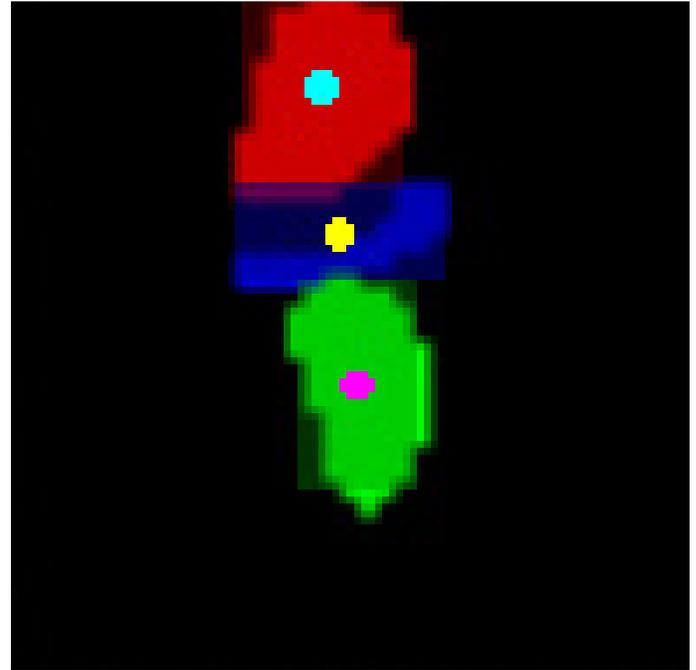

*Fig11 The segmentation with the center point and boxes*

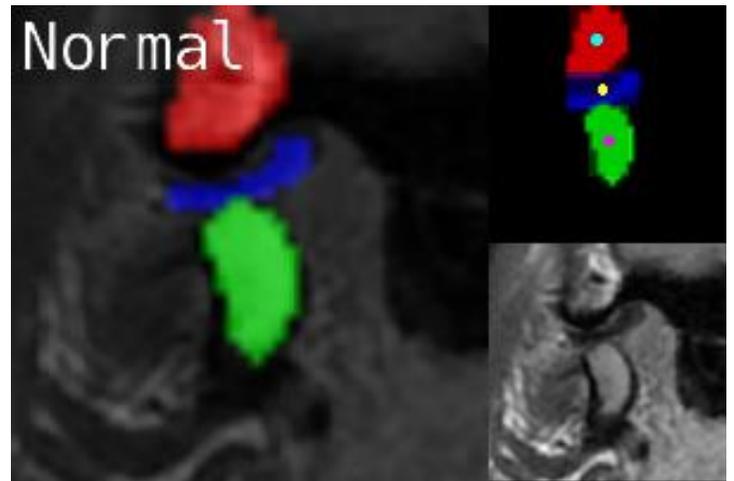

*Fig. 12 The overall picture. The left picture shows the overlaid segmentation of three labels on the MRI scan. The top right picture shows only the segmentation with center points and area boxes. The lower right picture shows the actual MRI scan.*

## VII. RESULT

In this study, we have evaluated the performance of the AI-driven TMD diagnostic system for diagnosing temporomandibular joint (TMJ) disorders. The results were compared with the ground truth labels provided by expert radiologists. Based on the classification results, we calculated various performance metrics, including the confusion matrix, sensitivity, specificity, accuracy, recall, and F1-score.

To evaluate the performance of the AI-driven TMD diagnostic system for diagnosing temporomandibular joint (TMJ)

disorders, a k-fold cross validation was used. k=10 where the entire dataset was randomized and 10 blocks with each containing 14 datasets were created [14].

For each trial, 8 blocks for the training set, 1 block for the validation set and 1 block for the testing set were used. By using the round-ribbon method of the blocks, each dataset had a chance to be tested. There were 10 trials called dir0 – dir9, were preformed. Each trial went through training and the created model based on this trial was used for the testing set [15].

Each test dataset comprising 14 images was used. The system was then applied to the testing set to identify and classify the TMJ structures as normal or abnormal. By using the trained model and custom python code, the testing images were classified.[16] The classification results were compared with the ground truth labels provided by expert radiologists to assess the performance of the AI model.

Below are the 14 testing images results of the trial named dir0.

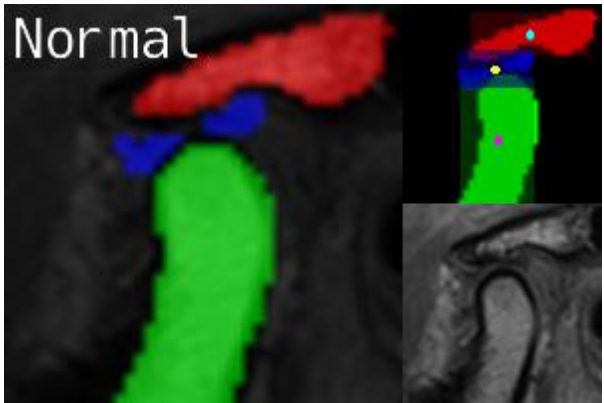
*Fig13-01 Correct diagnostic*

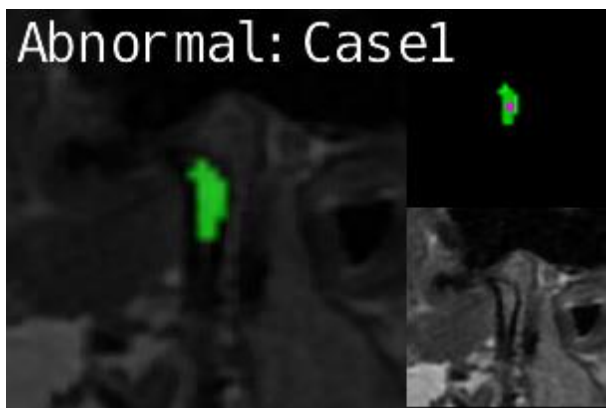
*Fig13-02 Correct diagnostic*

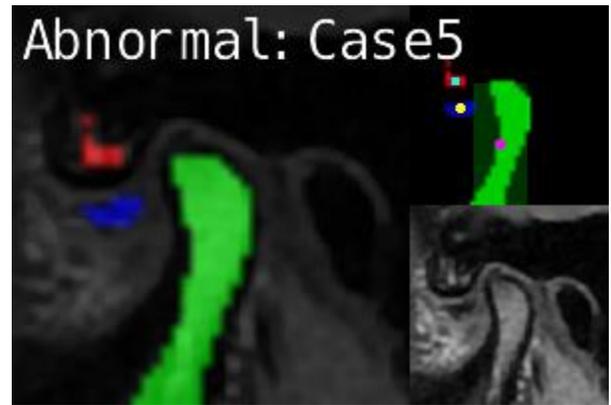
*Fig13-03 Correct diagnostic*

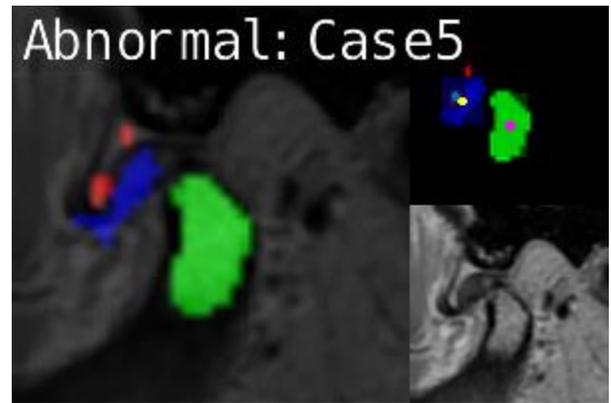
*Fig13-04 Correct diagnostic*

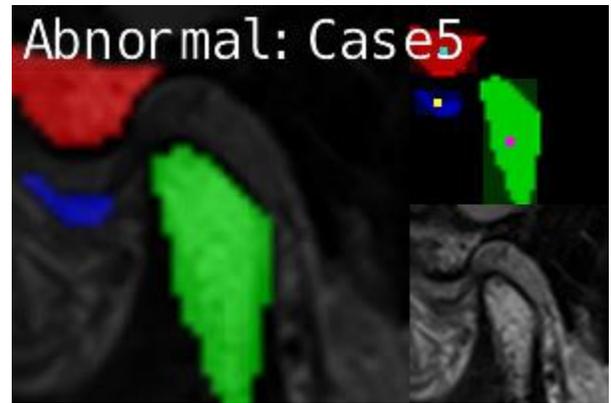
*Fig13-05 Correct diagnostic*

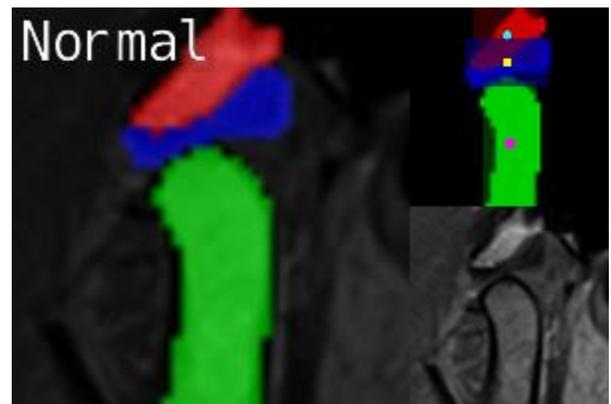
*Fig13-06 Correct diagnostic*

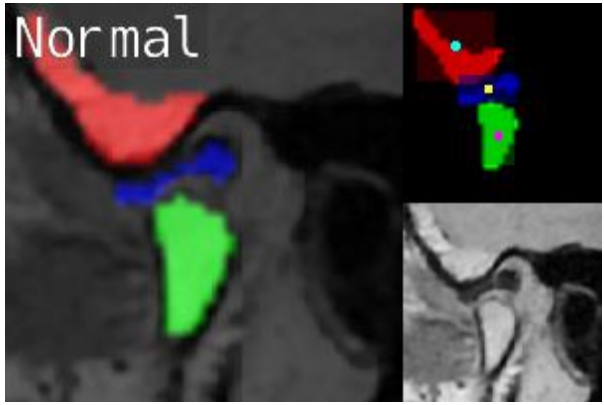

*Fig13-07 Correct diagnostic*

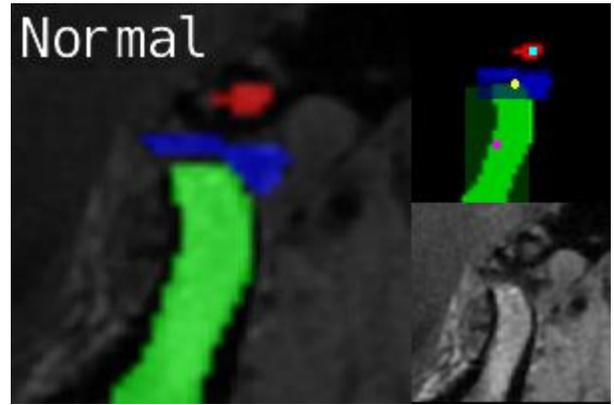

*Fig13-11 Correct diagnostic*

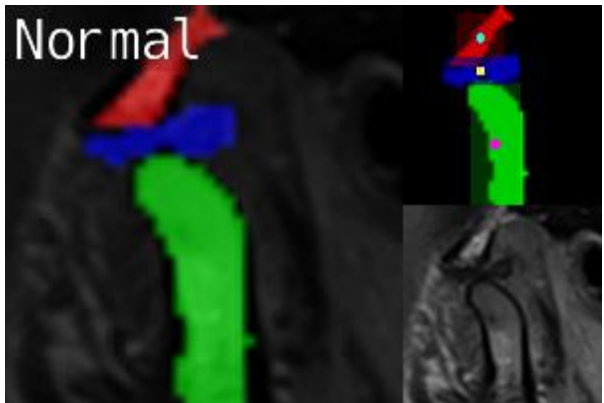

*Fig13-08 Correct diagnostic*

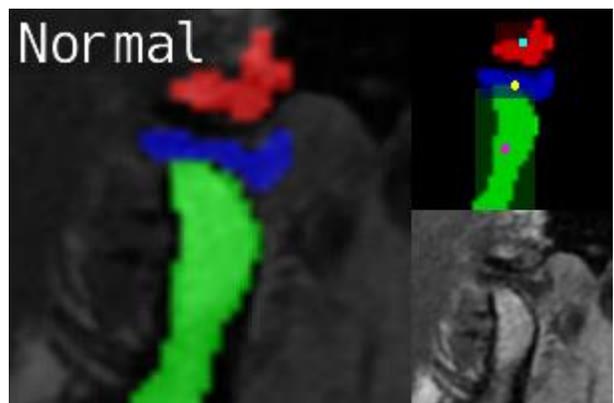

*Fig13-12 Correct diagnostic*

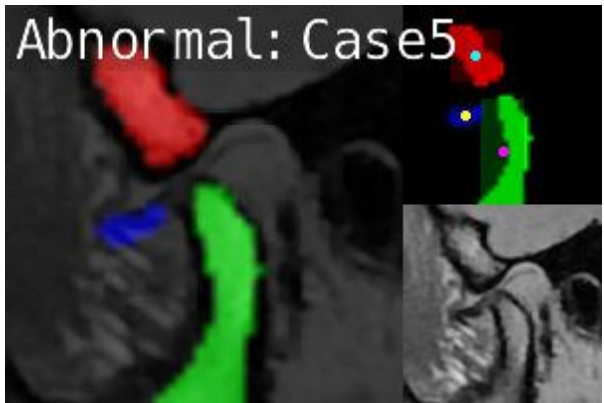

*Fig13-09 Correct diagnostic*

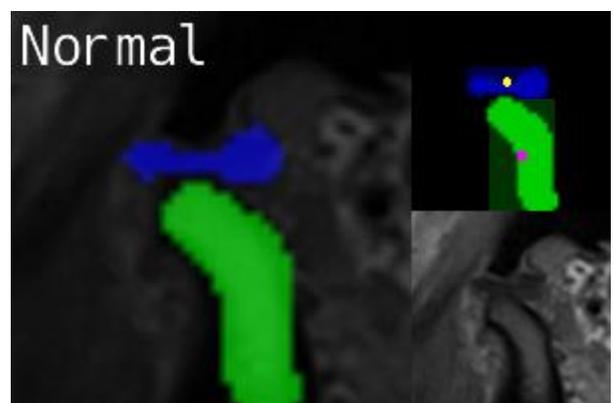

*Fig13-13 Correct diagnostic*

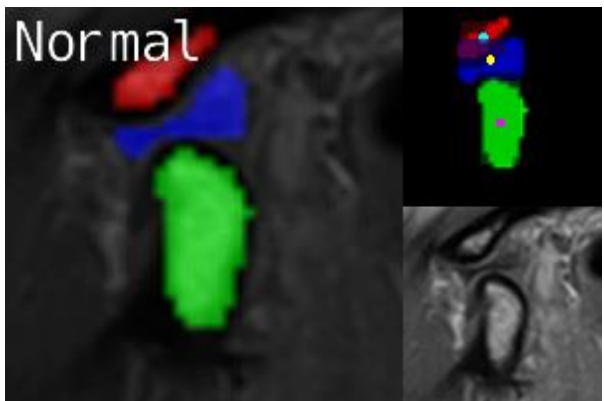

*Fig13-10 Correct diagnostic*

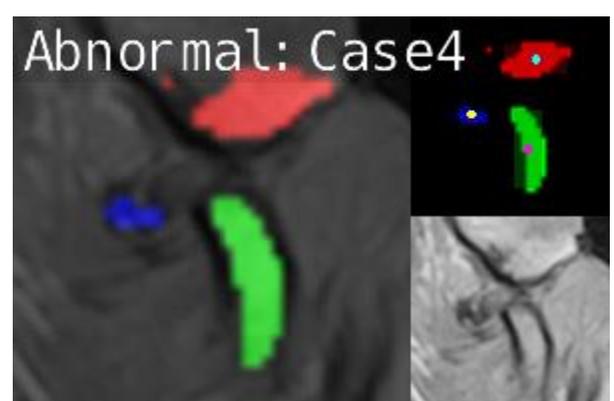

*Fig13-14 Correct diagnostic*

In Fig 14-1, The MRI image has incorrect diagnostic. It was the problem of segmentation. Specifically, the TMJ disc width was imporperly segmented (dir7, 8.jpg).

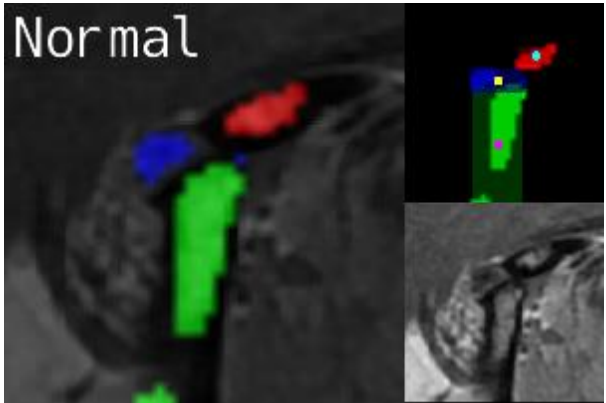

*Fig14-1 Incorrect diagnostic*

In Fig 14-2, The MRI image has also produced an incorrect diagnostic. It was the problem of segmentation. The TMJ disc labelled did not cover the top portion above the Condyle (dir9, 1.jpg).

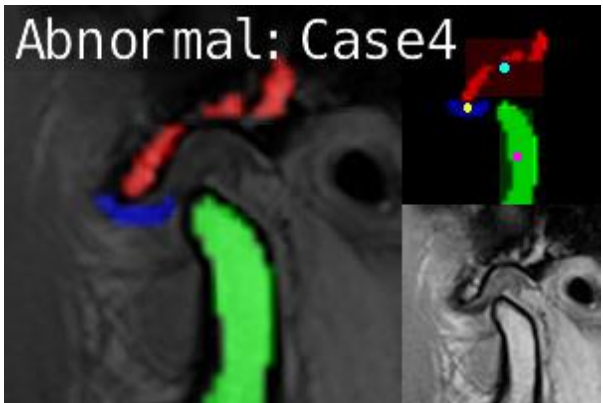

*Fig14-2 Wrong diagnostic*

## VIII. ANALYSIS

**The confusion matrix**
As there were 10 trials, all 10 results were presented in this section. In addition, the average scores for these 10 trials were calculated.

*TN (true negatives) represents the number of normal cases correctly identified.*
*FP (false positives) represents normal cases misclassified as abnormal.*
*FN (false negatives) represents abnormal cases misclassified as normal.*
*TP (true positives) represents the number of abnormal cases correctly identified.*

*Sensitivity: Sensitivity measures the proportion of actual abnormal cases that were correctly identified by the model.*
*Sensitivity = TP / (TP + FN)*

*Specificity: Specificity measures the proportion of actual normal cases that were correctly identified by the model.*
*Specificity = TN / (TN + FP)*

*Accuracy: Accuracy measures the proportion of correctly classified cases out of the total number of cases. Accuracy = (TP + TN) / (TP + TN + FP + FN)*

*Recall: In this context, recall is equivalent to sensitivity*

*Precision: Precision measures the proportion of true positive cases among the cases identified as positive by the model.*
*Precision = TP / (TP + FP)*

*F1-score: F1-score is the harmonic mean of precision and recall, providing a single metric that considers both false positives and false negatives. F1-score = 2 * (Precision * Recall) / (Precision + Recall)*

Below are the 10 trials.

| Predicted / Actual | Normal | Abnormal |
|---|---|---|
| Normal | TN=8 | FP=0 |
| Abnormal | FN=0 | TP=6 |

**Table 1-0.** The confusion matrix (dir0)
*From Table 1-0, dir0 has the following metrics:*
*Sensitivity(=Recall): 100%*
*Specificity: 100%*
*Accuracy: 100%*
*Precision: 100%*

| Predicted / Actual | Normal | Abnormal |
|---|---|---|
| Normal | TN=9 | FP=0 |
| Abnormal | FN=0 | TP=5 |

**Table 1-1.** The confusion matrix (dir1)
*From Table 1-1, dir1 has the following metrics:*
*Sensitivity(=Recall): 100%*
*Specificity: 100%*
*Accuracy: 100%*
*Precision: 100%*

| Predicted / Actual | Normal | Abnormal |
|---|---|---|
| Normal | TN=6 | FP=0 |
| Abnormal | FN=0 | TP=8 |

**Table 1-2.** The confusion matrix (dir2)
*From Table 1-2, dir2 has the following metrics:*
*Sensitivity(=Recall): 100%*
*Specificity: 100%*
*Accuracy: 100%*
*Precision: 100%*

| Predicted / Actual | Normal | Abnormal |
|---|---|---|
| Normal | TN=8 | FP=0 |

| Actual \ Predicted | Normal | Abnormal |
|---|---|---|
| Normal | | |
| Abnormal | FN=0 | TP=6 |

**Table 1-3.** The confusion matrix (dir3)

*From Table 1-3, dir3 has the following metrics:*
*Sensitivity(=Recall): 100%*
*Specificity: 100%*
*Accuracy: 100%*
*Precision: 100%*

| Actual \ Predicted | Normal | Abnormal |
|---|---|---|
| Normal | TN=7 | FP=0 |
| Abnormal | FN=0 | TP=7 |

**Table 1-4.** The confusion matrix (dir4)

*From Table 1-4, dir4 has the following metrics:*
*Sensitivity(=Recall): 100%*
*Specificity: 100%*
*Accuracy: 100%*
*Precision: 100%*

| Actual \ Predicted | Normal | Abnormal |
|---|---|---|
| Normal | TN=8 | FP=0 |
| Abnormal | FN=0 | TP=6 |

**Table 1-5.** The confusion matrix (dir5)

*From Table 1-5, dir5 has the following metrics:*
*Sensitivity(=Recall): 100%*
*Specificity: 100%*
*Accuracy: 100%*
*Precision: 100%*

| Actual \ Predicted | Normal | Abnormal |
|---|---|---|
| Normal | TN=9 | FP=0 |
| Abnormal | FN=0 | TP=5 |

**Table 1-6.** The confusion matrix (dir6)

*From Table 1-6, dir6 has the following metrics:*
*Sensitivity(=Recall): 100%*
*Specificity: 100%*
*Accuracy: 100%*
*Precision: 100%*

| Actual \ Predicted | Normal | Abnormal |
|---|---|---|
| Normal | TN=6 | FP=0 |
| Abnormal | FN=1 | TP=7 |

**Table 1-7.** The confusion matrix (dir7)

*From Table 1-7, dir7 has the following metrics:*
*Sensitivity(=Recall): 87.5%*
*Specificity: 100%*
*Accuracy: 92.9%*
*Precision: 100%*

| Actual \ Predicted | Normal | Abnormal |
|---|---|---|
| Normal | TN=10 | FP=0 |
| Abnormal | FN=0 | TP=4 |

**Table 1-8.** The confusion matrix (dir8)

*From Table 1-8, dir8 has the following metrics:*
*Sensitivity(=Recall): 100%*
*Specificity: 100%*
*Accuracy: 100%*
*Precision: 100%*

| Actual \ Predicted | Normal | Abnormal |
|---|---|---|
| Normal | TN=7 | FP=1 |
| Abnormal | FN=0 | TP=6 |

**Table 1-9.** The confusion matrix (dir9)

*From Table 1-9, dir9 has the following metrics:*
*Sensitivity(=Recall): 100%*
*Specificity: 87.5%*
*Accuracy: 92.9%*
*Precision: 85.7%*

Based on 10 trials:
*Average Sensitivity(=Average Recall): 98.8%*
*Average Specificity: 98.8%*
*Average Accuracy: 98.6%*
*Average Precision: 98.6%*
*Average F1-score (from Average Precision and Recall): 98.7%*

In summary, the AI-driven TMD diagnostic system demonstrated a sensitivity of 98.8%, a specificity of 98.8%, an accuracy of 98.6%, and an F1-score of 98.7% for diagnosing temporomandibular joint disorders. These results indicate that the model has high potential in assisting radiologists with TMJ diagnosis, but further improvements and validation using larger and more diverse datasets are required to enhance its performance and generalizability.

## IX. DISCUSSION

In this study, we have developed an AI-driven TMD diagnostic system using MRI images. These results indicate that this AI model has potential in assisting clinicians with TMJ diagnosis, but further improvements and validation using larger and more diverse datasets are needed to enhance its performance and generalizability.

The sensitivity of our AI system is relatively high (98.8%), suggesting that it can effectively identify abnormal TMJ cases. High sensitivity is crucial in medical diagnosis, as it helps to minimize the risk of human error, which could lead to delayed or missed diagnoses and subsequent complications. Balancing sensitivity and specificity is essential to optimize the performance of a diagnostic system and ensure that it does not produce too many false positives or false negatives.

One limitation of our study is the small sample size (140 images), which may have limited the model's ability to generalize to new cases. Previous studies have shown that deep learning models generally require large training datasets and various datasets to achieve optimal performance. Collecting and annotating a larger dataset, possibly through a multi-center collaboration, could help improve the model's performance and applicability to diverse populations.

Another potential limitation is the lack of a standardized MRI acquisition protocol, which could introduce variability in image quality and appearance. Standardizing the acquisition method and using data from multiple MRI machines and centers could help minimize potential errors and improve the model's performance.

Our study has contributed to the ongoing research regarding the use of AI in medical imaging. While several studies have investigated the use of AI for diagnosing TMJ disorders. Our study is one of the first to have specifically focused on segmentation methods. Our results suggest that AI-driven TMD diagnostic system has potential in this context, but further research is necessary to refine the approach and validate its performance in larger and more diverse populations.

## X. CONCLUSION

In conclusion, our AI-driven TMD diagnostic system has demonstrated potential in diagnosing temporomandibular joint (TMJ) disorders using MRI images. The model achieved relatively high sensitivity and acceptable specificity in detecting abnormal TMJ cases. However, the model's accuracy and F1-score indicate that room for improvement remains. Future studies should focus on utilizing advanced segmentation techniques, larger datasets, and standardized image acquisition protocols to enhance the performance and generalizability of the model.

Our study has contributed to the existing literature regarding the use of AI in medical imaging and has highlighted the potential of object detection methods in diagnosing TMJ disorders. The development of an AI-based diagnostic tool for TMJ disorders could help improve diagnostic accuracy and efficiency, reducing the burden on healthcare professionals and leading to better patient outcomes. With further refinement and validation, an AI-driven TMD diagnostic system could become an essential component of the diagnostic process for TMJ disorders and potentially other medical conditions as well.

## XI. REFERENCES

[1] D. Franklin, "Semantic Segmentation with SegNet." [Online]. Available: https://github.com/dusty-nv/jetson-inference/blob/master/docs/segnet-console-2.md.

[2] Z. Q. b. Guo Bai a 1, Qianyang Xie a 1, Hongyi Jing b, Shihui Chen c, Leilei Yu a, Zhiyuan Zhang a, Chi Yang a, "Automatic temporomandibular disc displacement diagnosis via deep learning," *Displays,* vol. 77, April 2023. [Online]. Available: https://www.sciencedirect.com/science/article/abs/pii/S0141938223000276.

[3] "TMD (Temporomandibular Disorders) from National Institue of Dential and Craniofacial Research," 2023. [Online]. Available: https://www.nidcr.nih.gov/health-info/tmd.

[4] R. Ohrbach, & Dworkin, S. F., "The Evolution of TMD Diagnosis: Past, Present, Future. Journal of Dental Research," pp. 1093–1101, 2016. [Online]. Available: https://doi.org/10.1177/0022034516653926.

[5] A. O. H Kurita 1, H Kobayashi, K Kurashina, "Resorption of the postero-superior corner of the lateral part of the mandibular condyle correlates with progressive TMJ internal derangement," 2003. [Online]. Available: https://pubmed.ncbi.nlm.nih.gov/14505617/.

[6] R. de Leeuw, & Klasser, G. D. (Eds.), "Orofacial Pain: Guidelines for Assessment, Diagnosis, and Management," 2013.

[7] D. Franklin, "Locating Objects with DetectNet." [Online]. Available: https://github.com/dusty-nv/jetson-inference/blob/master/docs/detectnet-console-2.md.

[8] J. L. Evan Shelhamer, Trevor Darrell, "Fully Convolutional Networks for Semantic Segmentation," May 20 2016. [Online]. Available: https://arxiv.org/abs/1605.06211.

[9] K. Wada, "Labelme." [Online]. Available: https://github.com/wkentaro/labelme.

[10] "Visual Object Classes Challenge 2012 (VOC2012)," 2012. [Online]. Available: http://host.robots.ox.ac.uk/pascal/VOC/voc2012/.

[11] "FCN_RESNET101." [Online]. Available: https://pytorch.org/vision/main/models/generated/torchvision.models.segmentation.fcn_resnet101.html.

[12] "Open Neural Network Exchange." [Online]. Available: https://onnx.ai/.

[13] J. Polikevičius, "pytorch-segmentation." [Online]. Available: https://github.com/Onixaz/pytorch-segmentation.

[14] AK51, "Split Code." [Online]. Available: https://github.com/AK51/pytorch-segmentation/blob/master/split_custom.py.

[15] AK51, "TMD code." [Online]. Available: https://github.com/AK51/pytorch-segmentation.

[16] AK51, "DecisionTreeCode." [Online]. Available: https://github.com/AK51/pytorch-segmentation/blob/master/segnet_GY_dot.py.